\documentclass[%
 reprint,
showpacs,showkeys,preprintnumbers,
 amsmath,amssymb,
 aps, prl
]{revtex4-1}

\usepackage{graphicx}% Include figure files
\usepackage{dcolumn}% Align table columns on decimal point
\usepackage{bm}% bold math

\begin{document}

\newcommand{\beq} {\begin{equation}}
\newcommand{\enq} {\end{equation}}
\newcommand{\ber} {\begin {eqnarray}}
\newcommand{\enr} {\end {eqnarray}}
\newcommand{\eq} {equation}
\newcommand{\eqs} {equations }
\newcommand{\mn}  {{\mu \nu}}
\newcommand{\sn}  {{\sigma \nu}}
\newcommand{\rhm}  {{\rho \mu}}
\newcommand{\sr}  {{\sigma \rho}}
\newcommand{\bh}  {{\bar h}}
\newcommand {\er}[1] {equation (\ref{#1}) }
\newcommand {\ern}[1] {equation (\ref{#1})}
\newcommand{\mbf} {{ }}

\preprint{APS/ChauncerPRL211011}

\title{Uncertainty Relation for Chaos}% Force line breaks with \\

\author{A. Yahalom}%
\email{asya@ariel.ac.il}
\author{M. Lewkowicz}
\email{lewkow@ariel.ac.il}
\author{J. Levitan}
\email{levitan@ariel.ac.il} \altaffiliation[Also at ]{Department of Physics, Technical University of Denmark, Lyngby 2800,Denmark}
\author{G. Elgressy}
\email{gilelg@hotmail.com}
\author{L.P. Horwitz}
\email{larry@post.tau.ac.il}
 \altaffiliation[Also at ]{School of Physics, Tel Aviv University, Tel Aviv 69978, Israel}
 \altaffiliation[And ]{Israel Institute for Advanced Research IYAR, Rehovoth, Israel}
\affiliation{%
 Ariel University Center of Samaria, Ariel 40700, Israel
}%

\author{Y. Ben-Zion}
\affiliation{
 Department of Physics, Bar Ilan University, Ramat Gan 52900, Israel
}%

\date{\today}

\begin{abstract}

A necessary condition for the emergence of chaos is given. It is
well known that the emergence of chaos requires a positive
exponent which entails diverging trajectories. Here we show that
this is not enough. An additional necessary condition for the
emergence of chaos in the region where the trajectory of the system goes through,
is that the product of the maximal positive exponent
times the duration in which the system configuration point stays in the unstable
region should exceed unity. We give a theoretical
analysis justifying this result and a few examples.

\end{abstract}

\pacs{ 05.45.Ac;~95.10.Fh;~05.45.Jn}

\keywords{Chaos, Lyapunov Exponent, Geometrical Analysis.}

\maketitle

There are many systems for which local instability leads to
chaotic behavior. It is often possible to characterize this local
instability in terms of the divergence of orbits detected by
computing Lyapunov exponents \cite{Gutzwiller}, or by geometrical
methods in cases where a geometric description becomes available,
such as in the application of the Jacobi metric and the time
dependence of the resulting Jacobi equation for geodesic deviation
\cite{Gutzwiller,PRE47,PRE48,PRL97,PRE96} for Hamiltonian systems,
or the criteria based on the curvature of the dynamical manifold
obtained from the recently developed geometric embedding method
(GEM) \cite{Horwitz}. These methods generally involve
the computation of an exponential divergence as an indication of
local instability. A given restricted region of
exponential divergence, however, may not result in significant
deviation of orbits, and therefore in chaotic behavior of the
system.  In this paper we give a quantitative bound which has the
form of an {\it uncertainty relation} characterizing the
effectiveness of the divergence in a locally unstable region. We
find a relation between the time of passage of the orbit through
an unstable region and the measure of
instability, for example, a negative eigenvalue in the Jacobi
equation, or the positive value of a Lyapunov exponent, which we denote in all these
cases by $\lambda$, which results in stability. This relation is of the
form of a product
\beq
 \Delta t \cdot |\lambda|_{max} > 1,
 \label{1}
 \enq
similar to that of an uncertainty relation between time and frequency
in optics, or in quantum theory. The basis for this relation is
that if there is an indication of exponential divergence, it is the time the
system spends in the region of instability which determines whether the
exponential divergence has reached a significant magnitude during
this period. In general, the coefficient $\lambda$ is time dependent,
but for a small region of instability, the relation \ern{1}
provides a useful measure. Moreover, this is a necessary condition of instability,
and if it is not satisfied by the trajectories under study, one cannot expect
chaotic behavior.
Formula (\ref{1}) is applied in several examples.

In a recent study \cite{Yahalom} of the restricted three body problem, involving
several configurations of an Earth, Sun, Jupiter type system, we
successfully applied the geometric criterion of \cite{Horwitz} based on a
geometrical embedding of the Hamiltonian orbits derived from a
conformal transformation of the original Hamiltonian of the form
\beq
 H = \frac{p^2 }{2 m_e} +  V(r, r_j, r_{12})
 \label{2}
 \enq
where
\beq
V(r, r_j, r_{12})= \frac{1}{ 2} m_j r_j^2 \omega_j^2- \frac{4\pi^2 m_e
}{ r} - \frac {4 \pi^2 m_j }{ r_j } - \frac{4\pi^2 m_e m_j }{r_{12}}.
\label{2a}
\enq
This potential has an adiabatic time dependence due to the
motion of the large planet (Jupiter) through the periodic dependence
(frequency $\omega_j$ of its radial variable $r_j$; $r$ is the radial
coordinate of the Earth from the Sun, and $r_{12}$ is the Earth-
Jupiter distance). Although the GEM criterion
was developed for time-independent potential systems, it was shown in \cite{Yahalom}
 that the time dependence in this problem was negligible,
provided the orbits did not pass too close to the boundaries of the
physical region.  Our prediction was that of stability with the
exception of some small regions of instability near the foci (apsides)
in highly
eccentric orbits.  These small regions of instability do not affect
the overall stability of the system. In contrast,
both Lyapunov methods and methods based on the time dependence of
frequencies appearing in the application of the Jacobi metric, which
predicted chaotic behavior for the system \cite{SafaaiSaadat}, indicated
instability over large regions of the orbits. One therefore must
understand why exposure of the orbits to small regions of instability
does not affect the outcome of the simulations, indicating completely
stable behavior.  These results were also obtained for the case of the
Jupiter mass going to zero, for which the GEM method predicts the
stability of the two-body Kepler problem correctly.

We have therefore investigated the application of the relation (\ref{1}),
 and found that it was indeed fulfilled in these cases. We
have furthermore applied the idea to other known potentially
 chaotic systems in dynamical regions (according to the choice of parameters)
where the passage time multiplied by the maximum
negative eigenvalue is small (but not zero) and found that these
systems remained stable in this regime.  Moreover, we found in these
examples that as the parameters are changed to induce
chaotic behavior, the degree of chaos was well correlated with the
growth of the uncertainty product. This is consistent with the interpretation
that the passage time for exponential deviations in the
orbits induced by the time
dependent eigenvalues of the Jacobi equation
are critical in the development of chaos for such Hamiltonian
systems. We have furthermore verified that a similar phenomenon
occurs for the standard Lyapunov analysis in problems for which it is
applicable. In the following we introduce a toy model for
this phenomenon, and present results from simulations
of motion based on a polynomial potential for the
Hamiltonian system describing the Toda problem. We conclude
 with a discussion of the Kepler problem.

Assuming that we have two trajectories with a difference
$\xi_i$, $i=1,2,3,...n$ in $n$ dimensions, we study a Jacobi equation with the simplifying
assumption that we are close enough to the tangent space such that the covariant
derivatives are well approximated by ordinary derivatives \cite{Gutzwiller}, i.e.,
 the geodesic deviation (Jacobi) equation is
\beq
 {\ddot \xi}_i = N_{ij}(t)\xi_j.
\label{3}
\enq
In the Lyapunov analysis:
\beq
 N^L_{ij} = -\frac{1}{m_i}\frac{\partial^2 V}{\partial x_i \partial x_j}
\label{Lya}
\enq
in which $x_i$ are the spatial degrees of freedom of the mechanical system, $V$ is
the potential of the mechanical system and $m_i$ are the masses of the particles.
In the GEM analysis \cite{Horwitz}:
\beq
 N^G_{ij} = -\frac{1}{m_i} \left( \frac{3}{2(E-V)}\frac{\partial V}{\partial x_i}\frac{\partial V}{\partial x_j} +
  \frac{\partial^2 V}{\partial x_i \partial x_j} \right)
\label{GEM}
\enq
For a slowly varying matrix $N$ with a positive eigenvalue, $\xi$
may appear to be exponentially divergent on a
small time interval.
If we follow the orbit for some time, the maximum positive
eigenvalue may go to zero, and the system will enter a stable
regime. If the eigenvalue tends to zero sufficiently rapidly, the
exponential divergence of the orbits, characterized by $\xi$, would not
be adequate to lead to chaos. To see how this phenomenon can develop
quantitatively, we compute the solution to the time
dependent \ern{3} in what  follows.

Let us define ${\dot\xi} = \eta_i$;  \ern{3} can then be
written
\beq
{\dot \zeta} = M \zeta,
\label{4}
\enq
in which
\beq
\zeta = \left(
\begin{array}{c}
\xi \\
\eta
\end{array} \right)
\label{zeta}
\enq
and
\beq
M = \left(
\begin{array}{cc}
0 & I \\
N & 0
\end{array} \right)
\label{M}
\enq
where $I$ is a $n \times n$ unit matrix. Obviously the value of $\zeta$
cannot diverge faster than what is dictated by the maximal positive eigenvalue of $M$
over a certain duration. Let us now look at a specific example.
For a $2\times 2$ (2D) $N$ we have a  $4\times 4$ $M$.
The eigenvalues $\lambda$ of $N$ are then determined by:
\beq
 \lambda^2 -\lambda {\rm Tr}N + \det{ N} =0
 \label{5}
 \enq
The eigenvalues $\mu$  of $M$ are determined by
\beq
 \mu^4 -\mu^2 {\rm Tr} N + \det{ N} =0.
 \label{6}
 \enq
so that $\mu^2 = \lambda$. Then, {\it local} stability is determined by
\ber
& & \lambda >0: \mu \ {\rm real, unstable}
\nonumber \\
& & \lambda <0, \mu \ {\rm imaginary, stable}
\label{7}
\enr
It might be convenient to take
\beq
N= a I + b \sigma_1 + c \sigma_3 \qquad (\sigma_i {\rm \ is \ a \ Pauli  \ matrix})
\label{8}
\enq
as a real symmetric example. Then,
\beq
N = \left(
\begin{array}{cc}
a+c & b \\
b & a-c
\end{array} \right),
\label{9}
\enq
and the eigenvalues become:
\beq
\lambda_{\pm} = a \pm \sqrt{b^2 + c^2}.
\label{10}
\enq
Now let us assume that:
\beq
b = \rho \cos{\theta}, \qquad
c = \rho \sin{ \theta},
\label{11}
\enq
where $\rho>0$ and $\theta$ are some arbitrary parameters. In this form the eigenvalues
will take the form:
\beq
\lambda_{\pm} = a \pm \rho.
\label{10b}
\enq
Notice that the value of $\theta$ does not effect the eigenvalues and hence
we will take it as $\theta=0$. Moreover by proper temporal scaling we can always take $\rho=1$.
For $a=-|a|<0$, the eigenvalue $\lambda_{-}$ is negative and hence always stable; we thus need
to analyze only the behavior of $\lambda_{+}$. The following form for $a$ is assumed:
\beq
a = -  \frac{t}{\Delta t}
\label{10c}
\enq
for which $\Delta t > 0$ is some given time interval. The eigenvalues are then:
\beq
\lambda_{\pm} = -  \frac{t}{\Delta t} \pm 1.
\label{10d}
\enq
In particular, $\lambda_{+} = 1 -  \frac{t}{\Delta t}$  is positive for $t < \Delta t$ (and hence unstable)
 but negative for $t > \Delta t$  and hence stable. This model is therefore
appropriate in order to study the effect of the duration of the trajectory in the unstable region on the systems
overall stability. The maximal instability exponent is $\lambda_{+max}=1$ at $t=0$.
Thus \ern{1} takes the simple form:
\beq
\Delta t > 1
\label{deltbig1}
\enq
as a necessary (but not sufficient condition) of instability. To appreciate the significance of
the above result let us look at two trajectories displaced at $t=0$ by an amount $\vec \xi = (1,0)$ but with
the same velocity $\vec \eta = (0,0)$. In the case $\Delta t = 5$ depicted in figure \ref{fig1}, it is clearly
seen that the exponential growth (black line) is not achieved by either component of $\vec \xi$. We
see a growth by an order of magnitude in the size of the displacement $\xi_1$, the displacement $\xi_2$ achieves
the same magnitude as $\xi_1$ although null at $t=0$ due to the coupling with $\xi_1$. The entire growth happens
at an interval slightly longer than $\Delta t = 5$ including a slight overshot. Entering the stable region the displacements
oscillate with a constant amplitude. Let us now consider the case in which \ern{deltbig1} is not satisfied,
for example let $\Delta t = 0.1$; this case is depicted in figure \ref{fig2}. One can hardly notice the exponential
growth in figure \ref{fig2}; as to $\xi_1$ and $\xi_2$ the fact that the "unstable" region is so minute does not allow
them to grow at all and they oscillate at an amplitude smaller than the original displacement. Having demonstrated the chaotic
uncertainty principle with a toy model we now move on to more realistic examples.
\figure
 \vspace{5cm}
 \includegraphics{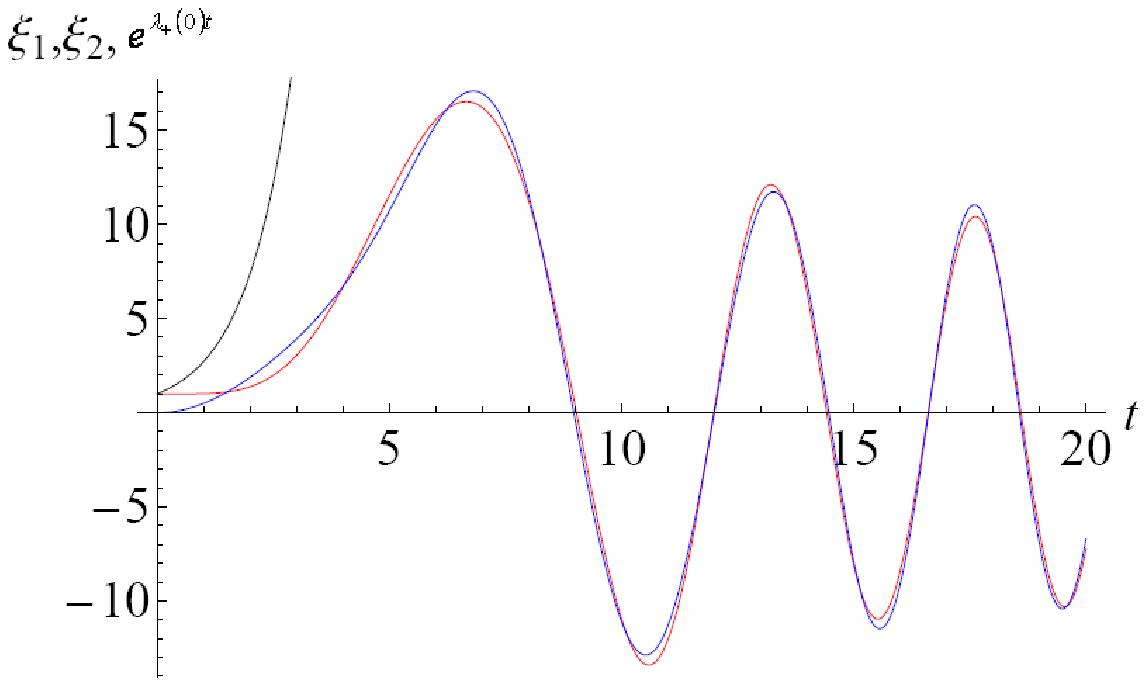}
\caption{The case $\Delta t = 5$, the figure depicts $\xi_1 (t)$ $(\xi_1 (0)=1)$ by a red line, $\xi_2 (t)$ $(\xi_2 (0)=0)$
by a blue line while the black line describes an exponential growth according to the maximal exponent $\lambda_{+max}=1$. (color online)}
\label{fig1}
\endfigure
\figure
 \vspace{5cm}
 \includegraphics{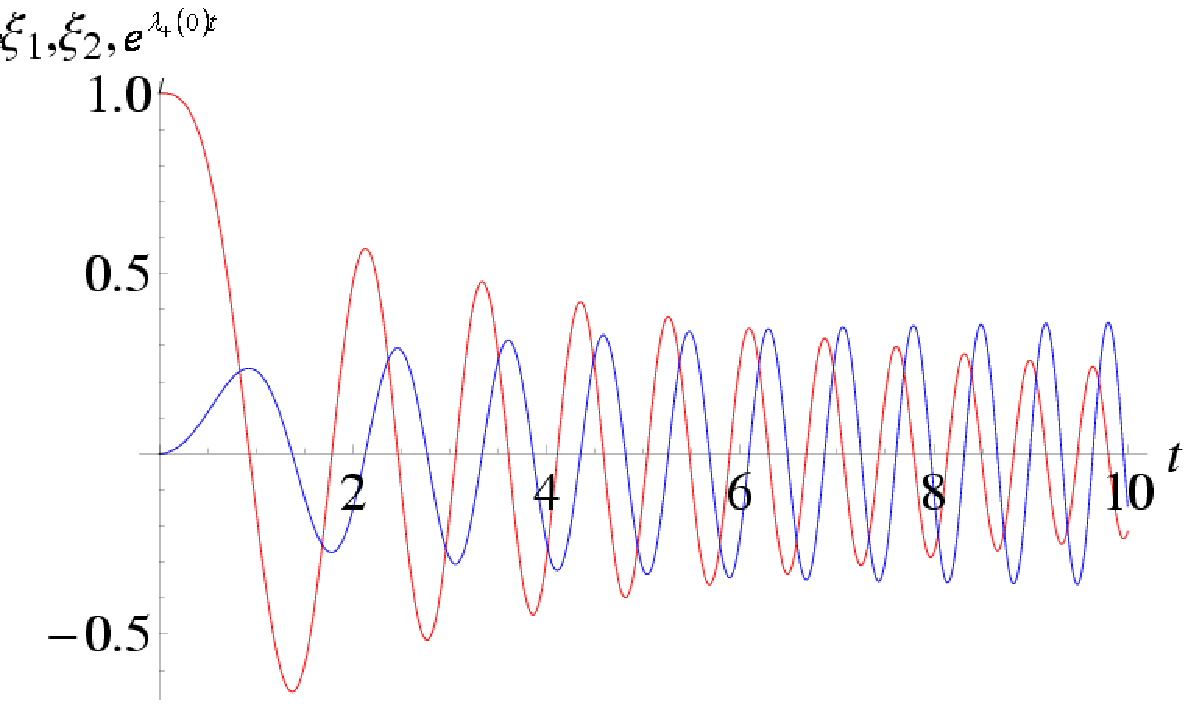}
\caption{The case $\Delta t = 0.1$, the figure depicts $\xi_1 (t)$ $(\xi_1 (0)=1)$ by a red line, $\xi_2 (t)$ $(\xi_2 (0)=0)$
by a blue line while the black line describes an exponential growth according to the maximal exponent $\lambda_{+max}=1$. (color online)}
\label{fig2}
\endfigure

A generalization of the Toda potential is described by \cite{BenZion}; in this model
the potential of the two-dimensional system is given by:
\beq
V(x,y)= \frac{1}{2}(x^2+y^2) + x^2 y - \frac{1}{3} y^3 + \frac{3}{2} x^4 + \frac{1}{2} y^4
\label{gToda}
\enq
Considering a unit mass, it was shown previously in \cite{BenZion} that for energy $E$ somewhat larger than $0.2$
the system becomes chaotic. Calculating the value of $\lambda \Delta t$ in which $\Delta t$
is the duration spent in unstable regions, one arrives at figure \ref{figToda}. It is clearly
\figure
 \vspace{7cm}
 \includegraphics{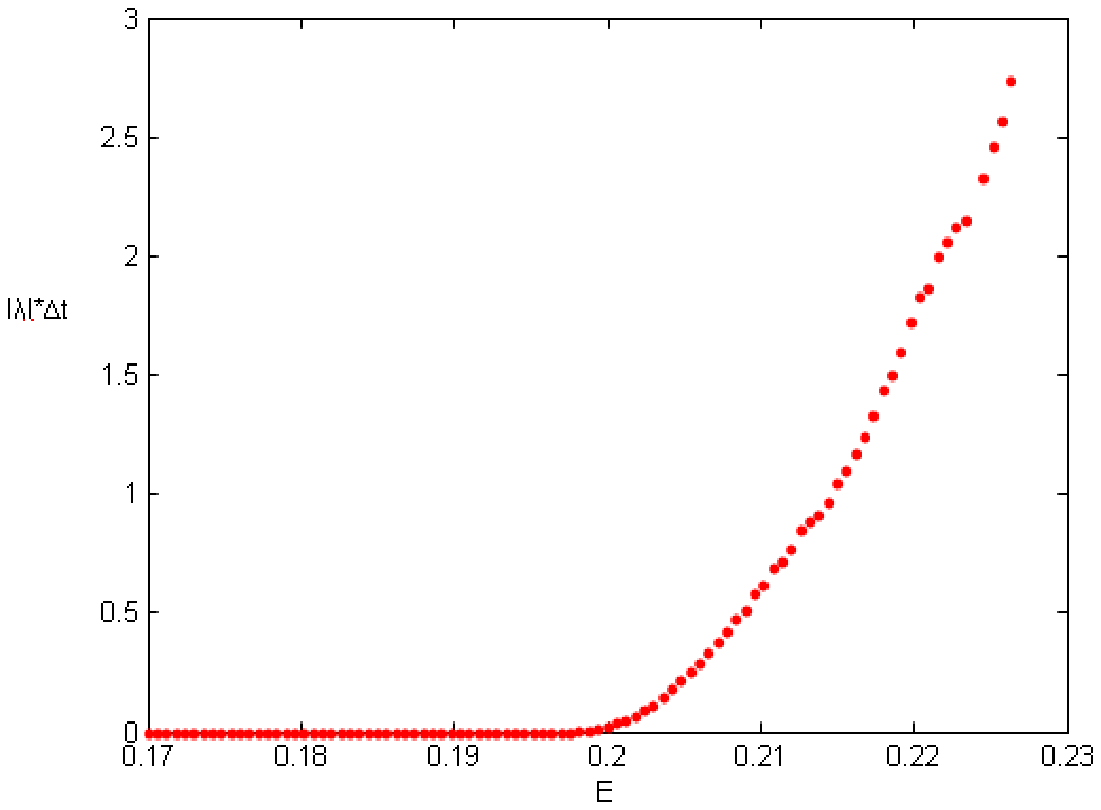}
\caption{The product of $\lambda$  with the time spent in unstable regions $\Delta t$ vs. energy. As the product becomes
appreciable, the increase of chaotic behavior is seen clearly on Poincar\'e plots.(color online)}
\label{figToda}
\endfigure
seen that the product of $\lambda$  with the time spent in unstable regions only reaches considerable size
when the system becomes chaotic, that is when $E > 0.215$. The Poincar\'e plots for $0.2 < E <0.215$  does not show chaotic behavior
 since the product of $\lambda$  with the time spent in the unstable region
is not considerable. This is shown in figure \ref{Poincare}. The Lyapunov exponent just below the chaotic threshold of $E=0.2$
(not shown here) predicts chaotic behavior which is not manifested; this is another indication of the effectiveness of the GEM method.
\figure
 \vspace{7cm}
 \includegraphics{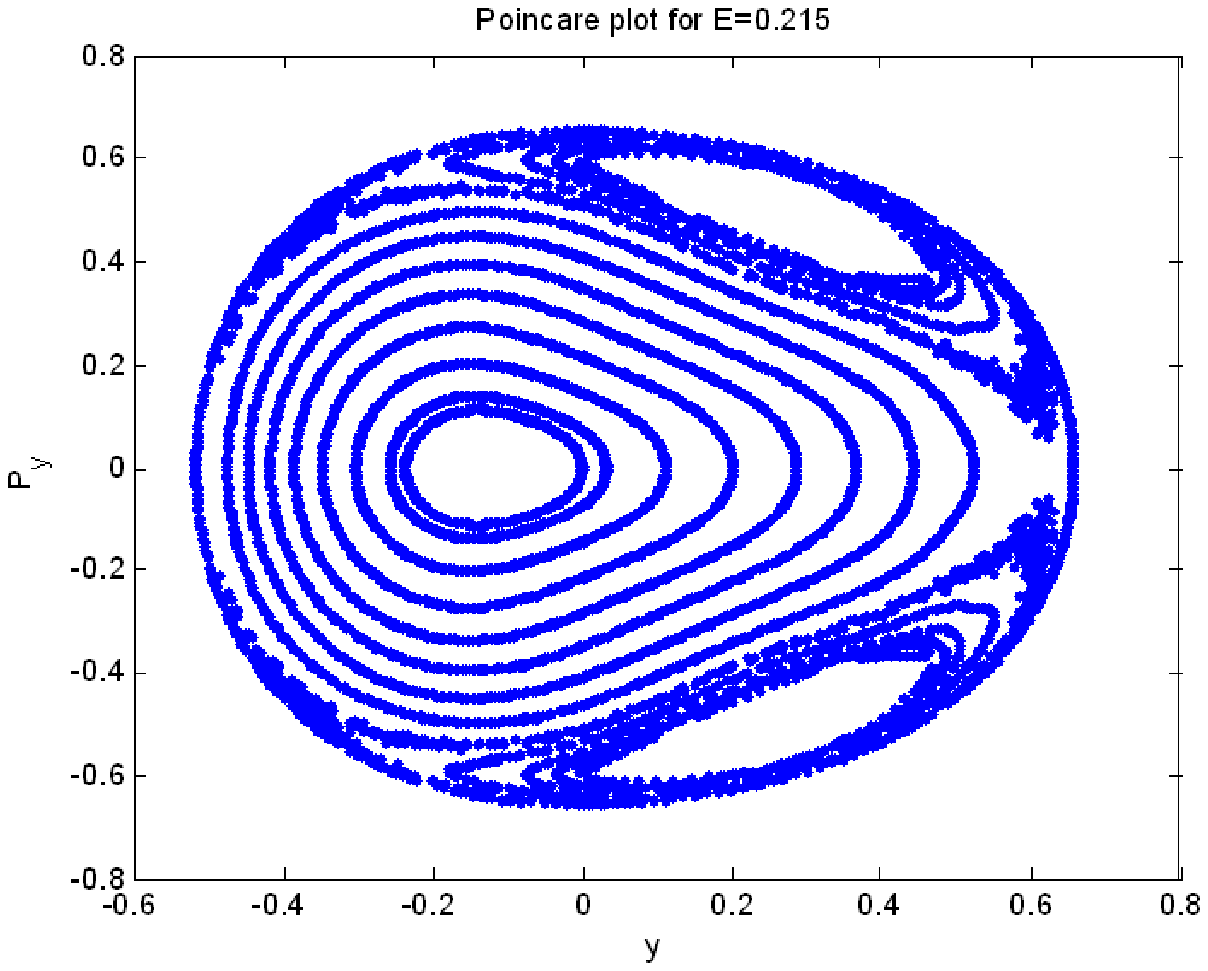}
\caption{The Poincar\'e plot for $E=0.215$. For this value $\lambda \Delta t = 1.044$
and a very mild chaotic behavior is seen as predicted by the chaotic uncertainty relation. (color online)}
\label{Poincare}
\endfigure

It is well known that the Kepler orbit which is clearly integrable and not chaotic,
is predicted to be chaotic according to the Lyapunov stability criterion. At least one Lyapunov exponent
is positive throughout the orbit indicating all over instability. In contrast, the geometric
method developed by Horwitz et al. \cite{Horwitz} generates the correct sign of the corresponding exponent.
The same is true for the restricted three-body problem in which a "Jupiter" is constrained
to move on a large circular trajectory \cite{Yahalom}. This system is described by equations (\ref{2}) and (\ref{2a}).
The above statement is true for low eccentricity orbits, however
for high eccentricity orbits such as the one depicted in figure \ref{eOrbit} even the GEM method suffers from
local unstable exponents near the perihelion. Nevertheless, a calculation of the product of the exponent times
the duration it takes the orbit to go through the unstable region indicates that the necessary
condition (\ref{1}) of instability is not met. For example the orbit of figure \ref{eOrbit} exhibits
\figure
 \vspace{2.5cm}
 \includegraphics{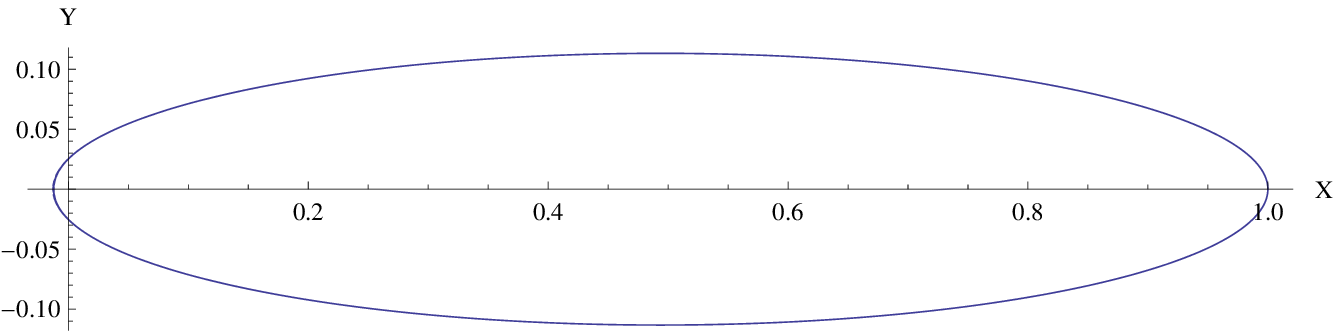}
\caption{A high eccentricity orbit. (color online)}
\label{eOrbit}
\endfigure
a positive geometric exponent near the perihelion (being stable everywhere else)
of the form described in figure \ref{lambeor}.
\figure
 \vspace{5cm}
 \includegraphics{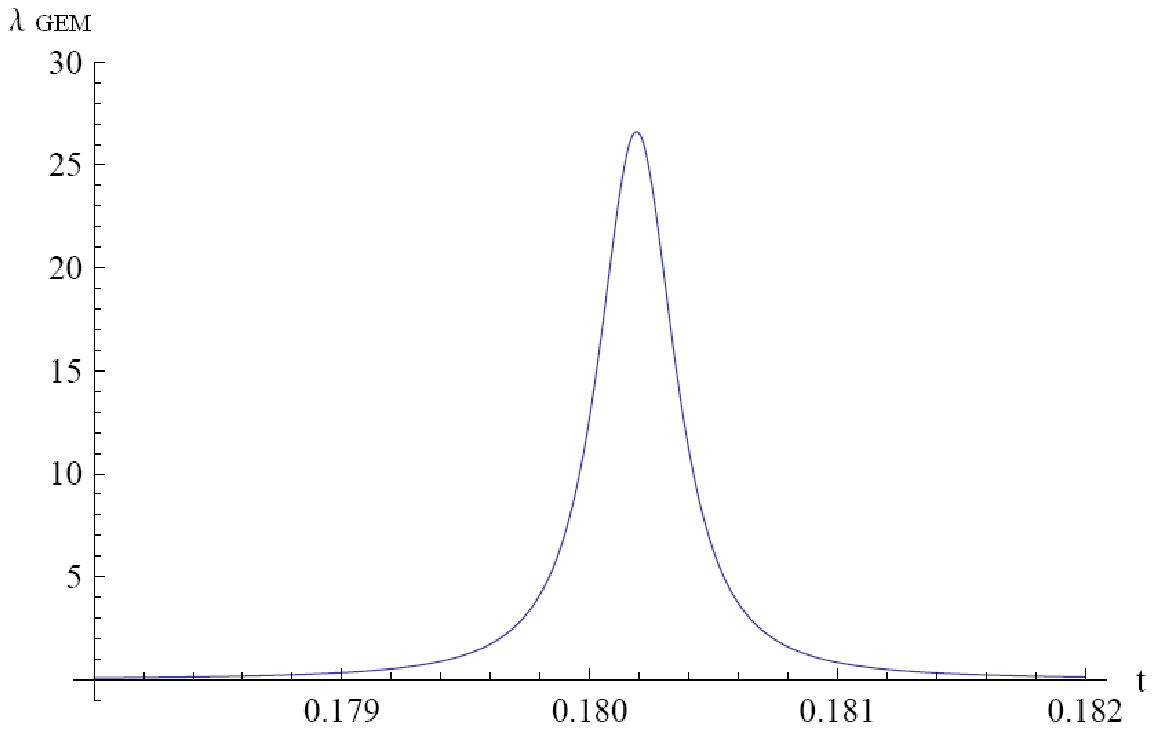}
\caption{The geometric eigenvalue of the Keplerian orbit near perihelion. (color online)}
\label{lambeor}
\endfigure
For this unstable region we have $\lambda_{max} \Delta t \simeq 0.6$, which is below what is required for chaotic instability.
The Lyapunov criterion, however, implies instability over the entire orbit.

Our analysis of the Kepler problem and the restricted three body problem using
the GEM method has led us to investigate the meaning of localized instabilities.
This in turn has led us to formulate the chaotic uncertainty principle of \ern{1}.
The content of the idea was investigated through a toy model demonstrating that
unless the condition is met one cannot expect small deviations to grow by more
than an order of magnitude. The value of this idea was further underlined in the generalized
Toda potential. Finally it proved most valuable in understanding the ineffectiveness
of the very local indications of instability arising in the Kepler and
three-body orbits that appear in the GEM analysis. We foresee
 the usefulness of the idea for studying chaotic systems in general.

\end {document}